\documentclass[%
reprint,
amsmath,amssymb,
aps,
]{revtex4-2}

\usepackage{physics}
\usepackage{graphicx}
\usepackage{dcolumn}
\usepackage{bm}
\usepackage{xcolor}
\definecolor{mycolor1}{rgb}{0.6350 0.0780 0.1840}
\usepackage{hyperref}
\hypersetup{
  breaklinks=true,
  colorlinks=true,
  allcolors=blue,
}

\begin{document}
	
	\preprint{APS/123-QED}
	
	\title{{Intertwining of} lasing and superradiance under spintronic pumping  } 
 
	\author{Oksana Chelpanova$^{1}$, Alessio Lerose$^2$, Shu Zhang$^{3,4}$, Iacopo Carusotto$^5$, Yaroslav Tserkovnyak${^{4}}$, Jamir Marino$^{1,6}$}
	\affiliation{%
	 $^1$Institut f\"ur Physik, Johannes Gutenberg Universit\"at Mainz, D-55099 Mainz, Germany \\
	$^2$Department of Theoretical Physics, University of Geneva, Quai Ernest-Ansermet 30, 1205 Geneva, Switzerland \\
	$^3$Max-Planck-Institut f\"ur Physik komplexer Systeme, 01187 Dresden, Germany \\
	$^4$Department of Physics and Astronomy, University of California, Los Angeles, CA 90095, USA\\
	$^5$INO-CNR BEC Center and Dipartimento di Fisica, Universit\`a di Trento, 38123 Trento, Italy\\
		$^6$Kavli Institute for Theoretical Physics, University of California, Santa Barbara, CA 93106, USA
	}%

	\date{\today}%

\begin{abstract} 
{We introduce a quantum optics platform featuring  the minimal ingredients for the description of a spintronically pumped magnon condensate, which we use to  promote driven-dissipative phase transitions in the context of spintronics. }
We consider a  {Dicke model
weakly coupled to an  out-of-equilibrium bath {with a tunable} 
spin accumulation. { The latter is pumped incoherently} in a fashion reminiscent of experiments with magnet-metal heterostructures.}
 {The core of our analysis is the emergence of}   
a hybrid lasing-superradiant regime that {does not take place  in an  ordinary pumped Dicke spin ensemble}, and { which can be traced back to   the spintronics pumping scheme}.
We interpret the resultant non-equilibrium phase diagram  from both a quantum optics and a spintronics standpoint, supplying a conceptual bridge between the two fields. 
The outreach of our results concern dynamical control
in {magnon condensates} and frequency-dependent gain media in quantum optics.
\end{abstract}

		\maketitle

	\textit{Introduction.}
	   The theme of dynamical phase transitions enabled by the interplay of  interactions, drive, and dissipation
{	   permeates different branches of quantum many body physics, such as }
	   quantum optics~\cite{carusotto2013quantum,PhysRevResearch.2.033131},  cold atoms~\cite{ritsch2013cold}, and non-equilibrium solid state physics~\cite{houck2012chip,RevModPhys.82.2731,PhysRevA.86.012116}. The interest in them ranges from practical applications  in dynamical control  to the fundamentals  of statistical mechanics. 
The exploration and understanding of non-equilibrium phases 
would benefit  from
	   a unifying 
	   language, 
	   which, however,
	   remains elusive
	   due to the diversity of
	   microscopic ingredients,  relevant scales, and engineering capabilities across the various platforms.

\begin{figure*}
    \centering
    \includegraphics[width=1\linewidth]{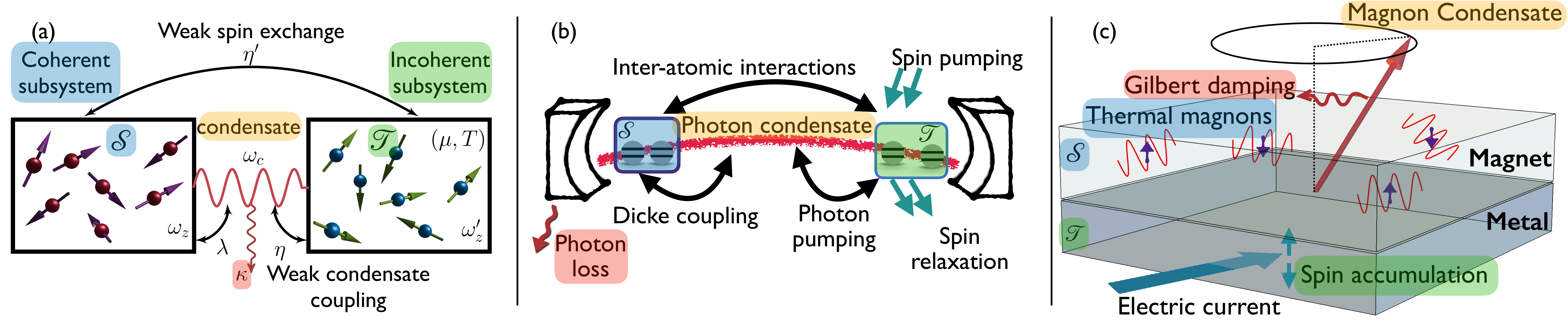}
    \caption{ 
    (a) 
    An ensemble of spins-1/2 ($\mathcal{S}$) is coupled to a  bosonic mode (in red) which  models a magnon condensate or a cavity boson which can become macroscopically occupied for large values of the Dicke coupling, $\lambda$.
	It is also coupled to a spin-1/2 subsystem ($\mathcal{T}$) under incoherent
	spin relaxation and pump. When $\mu>\omega_z^{\prime}$ the 
	population inversion
	of  $\mathcal{T}$
	can induce a coherent dynamical response in 
	$\mathcal{S}$, which  is the central mechanism  explored in this work. Dissipation with strength   $\kappa$ acts  on the bosonic mode  and 
			is shown by a red  wiggle line.  The model   contains the essential ingredients of both  quantum optics (b) and  spintronics (c) platforms,  as detailed in the main text. {Corresponding elements in different setups are highlighted in the same color.}}
    \label{fig:model}
\end{figure*}

In this Letter, we take a first step in filling this gap by studying
{a barebones model }
that offers complementary interpretations  
pertinent both to spintronics and driven-dissipative quantum optics, as illustrated in Fig.~\ref{fig:model}(a). 
We analyse a spin ensemble where  coherent   dynamical responses ascribable  to lasing can be induced by weak coupling to a subsystem  incoherently  pumped into a population inverted regime.  
%
{The model} can be realized 
{in}
an optical cavity, where two species of atoms, $\mathcal{S}$ and $\mathcal{T}$, are coupled to each other as well as to a common lossy cavity mode [cf. Fig.~\ref{fig:model}(b)].  
The ensemble $\mathcal{S}$ {collectively} couples to the cavity photon via a Dicke term, while the ensemble $\mathcal{T}$ does via a spin-boson interconversion term.
Dynamical instabilities can be induced 
{in the} subsystem $\mathcal{S}$ by incoherently spin pumping the subsystem $\mathcal{T}$, even in a limit of weak coupling
{between them.} 	
The interplay of the spin pumping and the Dicke coupling opens a parameter space where lasing and superradiant phases {intertwine} and lead to novel dynamical regimes exhibiting features of both.


\begin{figure}[]
\includegraphics[width=1\linewidth]{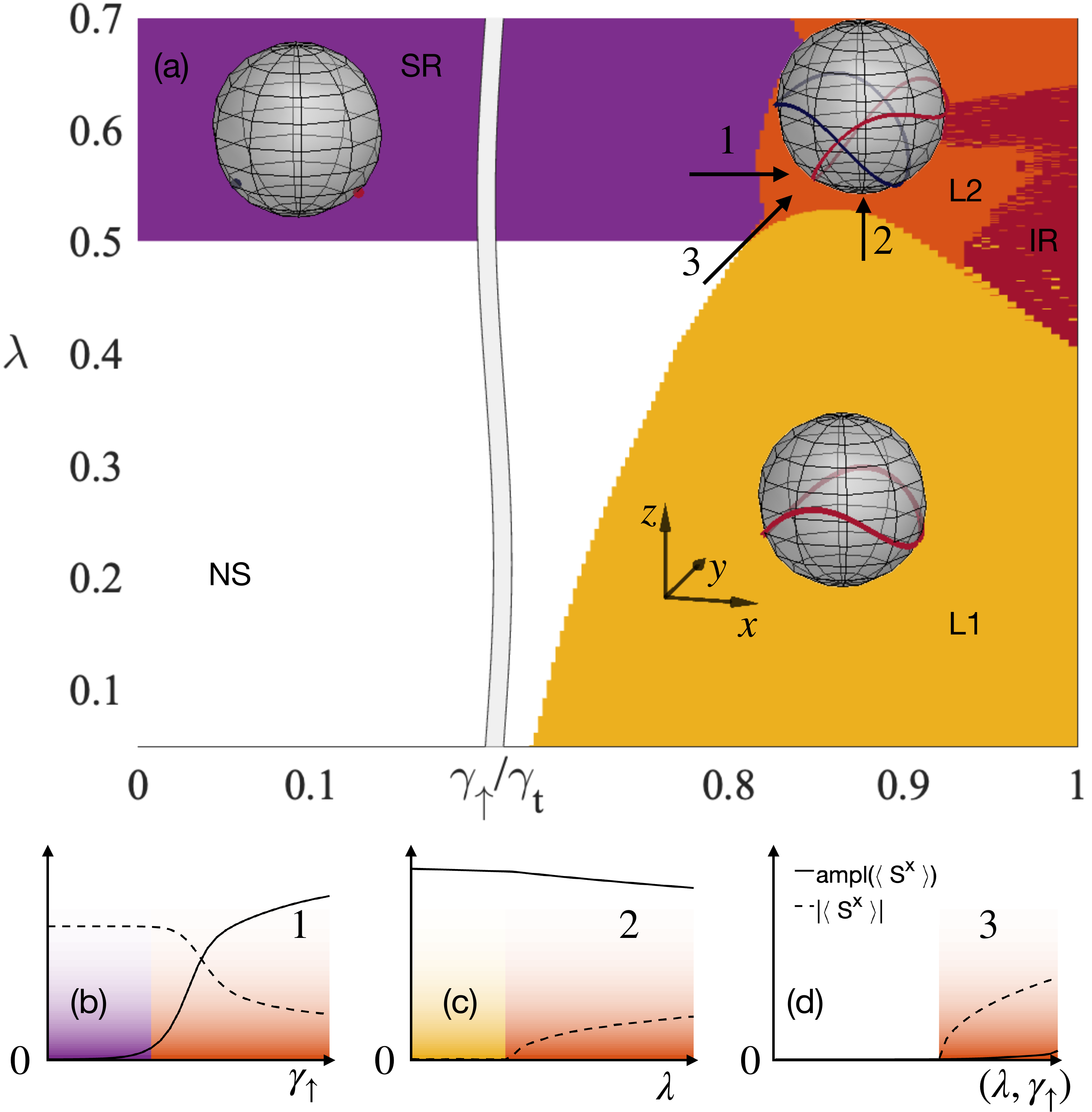}
	\caption{Dynamical phases resulting from the interplay of spin pumping and Dicke coupling: (a)  
	For $\gamma_{\uparrow}<\gamma_t/2$,   the usual critical coupling   ($\lambda_c\simeq 0.5$) associated to   the Dicke transition, separates the normal (NS) from the superradiant (SR) phase.  For $\gamma_{\uparrow }>\gamma_t/2$, the normal state   becomes unstable, and observables in the ensemble $\mathcal{S}$   oscillate  with zero average value of $\mathcal{S}^x$ in the  region L1,  and around one of   minima of the SR phase in the region L2, as shown in the Bloch spheres.   Inside the   irregular (IR) region the motion of the collective spin covers uniformly {a large part of} the Bloch sphere without   any structured pattern, and suggestive of  chaotic behaviour. 
  Panels (b)--(d) show the absolute value of  time-averaged  $\langle \mathcal{S}^x \rangle$ (dashed line) and amplitude of its oscillations (solid line) along the transition lines 1, 2 and 3 in the main inset (a).
  Here we have chosen $\omega_c=\omega_z=\omega_z^{\prime}=1,$ $\gamma_t=1$, $\eta=\eta^{\prime}=0.1,$ $\kappa=0.06.$}
	\label{fig:drawingfigureforpaper}
\end{figure}

 Our motivation to separate the coherent ($\mathcal{S}$) and incoherent ($\mathcal{T}$) spin subsystems stems from a solid-state viewpoint, to allow quantum correlations to settle in without much disruption from direct pumping processes. Considering magnet-metal heterostructures~\cite{Hauser1969,saitoh2006,Saitoh2010,Saitoh2010-2,Goennenwein2011,Chien2012,Goennenwein2013,Shi2013,Youssef2013} as 
{a primary} 
 example,
 the magnet layer has a stiff order parameter accompanied by coherent excitations ~\cite{kittel1963},
  while itinerant electrons carrying incoherent spins in the metal layer 
  are more amenable to external control~\cite{dyakonov1971}.
 One of the consequences of the magnet being a strongly interacting system is the propensity of  a long-wavelength magnon to undergo (Bose-Einstein) condensation~\cite{Slavin2006,Slavin2008,Duine2017book}, which is mimicked by 
 the   bosonic mode in our model. 
 In a magnet, such condensation can manifest as a static phase transition~\cite{giamarchi2008}, or a dynamical one with the magnetic order parameter precessing spontaneously~\cite{Berger1996,Volovik2007,Volovik2008}, bearing analogy to the superradiant and lasing transitions, respectively. 
 As shown in Fig.~\ref{fig:model}(c), a magnon condensation can be triggered by electrically pumping the heterostructure~\cite{Bender2012,Bender2014,Brataas2017,Takei2019,Huebl2019}.
  A spin accumulation is induced via the spin Hall effect in the metal~\cite{hurd2012,Hirsch1999,kato2004,sih2005,wunderlich2005,Tinkham2006} and exerts a spin torque~\cite{Slonczewski1996,Stiles2008,Tserkovnyak2014} on the magnetic dynamics {by interfacial magnon-electron scatterings}. {Such a torque can overcome the intrinsic magnon decay and maintain a quasi-equilibrium condensate of magnons.}
In addition, the magnon condensate
and the thermally occupied short-wavelength magnons 
undergo coupled dynamics,
previously described by a two-fluid theory~\cite{Flebus2016}.
Our model, though much simplified from this practical scenario, allows for a full   treatment of the interplay of spin pumping, coupling between the interacting magnetic  system and  pumped reservoir, and  dissipative effects. 
{We argue that the emergence of a dynamical phase, intertwining lasing and superradiance as a result of a pumping scheme inspired from spintronics (phase L2 in Fig.~\ref{fig:drawingfigureforpaper}), and yet   realizable in a quantum many body optics platform, can provide   a conceptual bridge between the two communities. }\\ 

\textit{Model.} We consider a 
Dicke sample~\cite{kirton2019introduction,EmaryBrandes,bhaseen2012dynamics,keeling2010collective,reiter2020cooperative},
{which consists of an ensemble $\mathcal{S}$ of $N$ spins-1/2 collectively coupled to a bosonic mode $a$ of frequency $\omega_c$,}
weakly interacting with 
{an ensemble $\mathcal{T}$ of an additional set of $N$ spins.}
The level splitting of spins in subsystem $\mathcal{S}$ ($\mathcal{T}$) is $\omega_z$ ($\omega_z^{\prime}$).
The full Hamiltonian reads  
	\begin{widetext}
		\begin{equation}\label{eq:main_big}
			H=\omega_c a^{\dag}a
			+\omega_z \mathcal{S}^{z}
			+\omega_z^{\prime}\mathcal{T}^{  z}
			+\frac{\lambda}{\sqrt{N}}\left(a+a^{\dag}\right)\left(\mathcal{S}^{+}+\mathcal{S}^{-}\right)
			+\frac{\eta}{\sqrt{N}}\left(a \mathcal{T}^{  +}+a^{\dag}\mathcal{T}^{  -} \right)
			+\frac{\eta^{\prime}}{N}\left(\mathcal{S}^{+}\mathcal{T}^{ -}+\mathcal{S}^{-}\mathcal{T}^{ +} \right),
		\end{equation}
	\end{widetext}
	\noindent where $a$ and $a^\dag$ are  bosonic annihilation/creation operators, mimicking the magnon condensate or the cavity photon, while the collective spin operators
	are  $\mathcal{S}^{-}=\sum_{i=1}^N\sigma_i^-$ and $\mathcal{T}^{  -}=\sum_{i=1}^N \tau_{i}^{-}.$  
	Here $\sigma^\alpha_i$  and $\tau^\alpha_i$ with $\alpha=x, y, z$ are 
{spin-1/2 operators}. We have introduced the Dicke coupling $\lambda$, a small boson-spin interconversion term $\eta$, and a small spin exchange coupling $\eta^{\prime}$.
	
The ensemble $\mathcal{T}$ is driven incoherently into a grand-canonical state
{with}
temperature $T$ and spin accumulation $\mu$, 
{by}
spin pump with rate $\gamma_\uparrow$ and loss
$\gamma_\downarrow$, 
which is described by the following Lindblad master equation~\cite{breuer2002theory} for the joint density matrix  of the total system:
	\begin{equation}\label{eq:master}
		\frac{d \rho}{d t}=-i\left[H, \rho\right]+\kappa \mathcal{D}[a]+\gamma_{\uparrow}\sum_{i=1}^{N} \mathcal{D}[ \tau_i^{+}]
		+\gamma_{\downarrow}\sum_{i=1}^{N} \mathcal{D}[\tau_i^{-}],
	\end{equation}
	neglecting spin dephasing effects~\cite{kirton2019introduction}.
	The dissipators
	$\mathcal{D}[{x}]\equiv {x} \rho {x}^{\dagger}-{1}/{2}\left\{ {x}^{\dagger} {x},\rho  \right\}$  are defined as usual, and the spin pump and loss rates
$\gamma_{\uparrow}={\gamma_t}/(1+e^{\beta\left(\omega_{z}^{\prime}-\mu\right)})$ and $\gamma_{\downarrow}={\gamma_t}/(1+e^{-\beta\left(\omega_{z}^{\prime}-\mu\right)})$ are {parametrized} by $\beta \!=\! T^{-1} \!>\! 0$ and $
\mu$, with $\gamma_t=\gamma_{\uparrow}+\gamma_{\downarrow} \geq 0$.
	{Evolution according to}
	Eq.~\eqref{eq:master}  drives the system into a mixed state with a relative population of up and down spins controlled by the ratio  $\gamma_{\uparrow}/\gamma_{t}$.
When $\mu>\omega'_z$, 
the incoherent subsystem $\mathcal{T}$  
experiences population inversion
which can be transferred to the rest of the system  via  $\eta$ and $\eta'$ and trigger a lasing instability. 
For $\gamma_t\gg \eta, \eta^{\prime}$,
	it 
	quickly relaxes towards a steady state with 
$\langle \mathcal{T}^{  z}\rangle  \approx (\gamma_{\uparrow}-\gamma_{\downarrow})/{2\gamma_t},$ and   $\langle \mathcal{T}^{  \pm}\rangle \approx 0.$

 In the dissipative dynamics of Eq.~\eqref{eq:master},
 we have also considered photon loss with rate $\kappa$ in order to model the photon line-width of the cavity. 
The relaxation of the collective bosonic mode in a magnet, on the other hand, depends self-consistently on its dynamics~\cite{mayergoyz2009}. {We therefore consider, as alternative, a viscous damping of the magnon condensate, whenever the spintronic relevance is concerned. 
In terms of magnetic dynamics, the phenomenological Gilbert damping~\cite{Gilbert2004} slows down the coherent precession of the order parameter and brings it towards the global equilibrium state~\cite{Stiles2008,vonOppen2011}.} 
Interestingly, our results remain qualitatively unaltered
{under} dissipation through photon loss or Gilbert damping~\cite{suppmat}. 

Before moving to a thorough  discussion of our results, we 
{remark}
that the model in Fig.~\ref{fig:model} should not be regarded as a faithful  modelization of an actual spintronics system. For instance, the Dicke coupling does not naturally occur in {magnets}.
Rather, our model contains the key ingredients for interplay of coherent interactions, spin pumping and magnon damping in a spintronics platform,
{to reveal} the mechanisms for the formation of novel dynamical phases which could then be  explored in the future within  realistic  devices.  

\emph{Superradiance and lasing. } 
{In presenting results below we use normalized variables $a\propto a/\sqrt{N},$ $S\propto S/N,$ $\mathcal{T}\propto \mathcal{T}/N$, as customary in the treatment of systems with collective light-matter interactions~\cite{kirton2019introduction,bhaseen2012dynamics,EmaryBrandes}.}
We start  by revisiting some established dynamical regimes  of the hamiltonian in {Eq.}~\eqref{eq:main_big}. 
	For   $\eta=\eta^{\prime}=0$, we recover a standard Dicke model~\cite{dicke1954coherence,bhaseen2012dynamics}. With $\lambda<\lambda_c=\sqrt{(\omega_z(\omega_c^2+\kappa^2/4))/{4\omega_c}}$  the system is in the normal state (NS) with a vanishing $\langle \mathcal{S}^x\rangle$ 
	component  and
	{ no macroscopic occupation of the photonic mode.}
 By increasing $\lambda>\lambda_c$ the system enters a   super-radiant  (SR) phase where it spontaneously breaks $\mathbb{Z}_{2}$ symmetry,   exhibiting $\langle \mathcal{S}^x \rangle \ne 0$ and   photon  condensation, $n=\langle a^\dag a \rangle\neq0$.
This picture 
{remains valid when} small $\eta$ and $\eta'$ are switched on
{while the} spin pumping is kept weak, namely   $\gamma_{\uparrow}<\gamma_{t}/2$ (cf.  Fig.~\ref{fig:drawingfigureforpaper}(a)).

{Another} limit  $\eta^{\prime}=\lambda=0$  corresponds to the incoherently pumped Tavis-Cummings  model~\cite{Kirton_2018,kirton2017suppressing,tieri2017theory, kirton2019introduction,bhaseen2012dynamics,kopylov2015dissipative}. 
The choice   trivializes the dynamics of the ensemble $\mathcal{S}$.
For 
${\gamma_{\uparrow}}/{\gamma_t}\ge {1}/{2}+{\kappa\gamma_t}/(8\eta^2)
+{\kappa\gamma_t(\omega_c-\omega_z^{\prime})^2}/(2\eta^2(\kappa+\gamma_t)^2)$,  
the $\mathcal{T}$ spins  
experience population inversion, with $\langle a\rangle$ and $\langle \mathcal{T}^{ x,y}\rangle$ undergoing oscillations. At long times,  both $\langle \mathcal{T}^z\rangle$ and the photon number  approach the {steady} values set by the pumping rates $\gamma_{\uparrow/\downarrow}$~\cite{suppmat}.

	\textit{Dynamical phase diagram. } By turning on  $\lambda$ together with sizeable spin pumping
{in}
	a weakly coupling limit ($\eta$, $\eta' \to 0^+$),  we generate 	the   diagram of   dynamical responses      (cf.  Fig.~\ref{fig:drawingfigureforpaper})   in  mean-field  treatment, which is exact for $N\to\infty$~\cite{bhaseen2012dynamics,dalla2013keldysh,lang2016critical}.
	In the Supplemental Materials~\cite{suppmat} we present the associated equations of motion and
	also analyze the breakdown of mean field from finite $N$ corrections. 

	For strong pumping ($\gamma_{\uparrow}>\gamma_{t}/2$), the spins in the ensemble $\mathcal{S}$ display long-lived oscillatory dynamics (see Bloch spheres in Fig.~\ref{fig:drawingfigureforpaper}).
The region L1 in Fig.~\ref{fig:drawingfigureforpaper} resembles the  regular lasing~\cite{kirton2017suppressing} discussed above, while L2 { features   'supperradiant'  oscillations}. 
The transition from L1 to L2 occurs around values of the Dicke coupling~$\sim \lambda_c$, with a non-vanishing time average  of $\langle \mathcal{S}^x\rangle $ in L2. In this  phase we observe   persistent oscillatory dynamics reminiscent of lasing around one of the symmetry-broken states of the Dicke model. Such 'superradiant' oscillations would not arise by directly pumping a Dicke model through the  Lindblad channels in Eq.~\eqref{eq:master}; they are a result of the pumping scheme of Fig.~\ref{fig:model} conceptually borrowed from spintronics. In this regard, the dynamical phase L2 is a conceptual 'bridge' between the quantum optics and spintronics communities which we are aiming to lay out in this work. 
{Notice that despite the pumped subsystem experiences population inversion, the spin ensemble $\mathcal{S}$  remains in a state with  negative $\langle \mathcal{S}^{z} \rangle$ in both phases L1 and L2.}

We now discuss the role of symmetries in the oscillatory dynamics displayed in L1 and L2, and in the transitions between these two different regimes.
	 For $\lambda=0$ the photon number $n$ does not oscillate. 
A nonzero $\lambda$ breaks
	 the U(1) symmetry
	 and  the oscillations in $n$ can be attributed to  ellipticity (i.e., different amplitudes of oscillations of $\langle \mathcal{S}\rangle$ spin components along $x$ and $y$ directions due to the presence of Dicke-like interaction term) in the spontaneous procession in   absence of $\mathcal{S}^z$ conservation. In fact, the dynamics are instead governed by a $\mathbb{Z}_{2}$ symmetry, reflected in the observation that  the oscillatory frequency of $n$ and $\langle \mathcal{S}^{z} \rangle $ is twice that of $\langle \mathcal{S}^x \rangle $.
	 The transition from   L1 to L2  is characterized by an increase in the  time-average value of $\langle \mathcal{S}^x \rangle$, 
    which
	 can be explained by the 
	 {spontaneous}
	 breaking of the $\mathbb{Z}_{2}$ symmetry upon increasing the Dicke coupling   $\lambda$ [cf. Fig.~\ref{fig:drawingfigureforpaper}(c)].  
	 {The transition from the SR region to the L2  region appears as a crossover in finite-time numerical data, as the damping of the oscillations of $\langle \mathcal{S}^x \rangle$ critically slows down upon approaching the transition point from the SR side, hence   the time-averaged  amplitude of the oscillations in a long but finite time windows smoothly   grows, blurring
	 the expected singular behavior at the phase boundary} [cf. Fig.~\ref{fig:drawingfigureforpaper}(b)] 
associated to the dynamical     spontaneous symmetry breaking of the U(1) symmetry.  
	 Finally, in the transition from NS to L2,  the
	 absolute value of the time average of  $\langle \mathcal{S}^x \rangle $, as well as its amplitude, build up [cf. Fig.~\ref{fig:drawingfigureforpaper}(d)].

This dynamical phase diagram with competing {stationary} (NS, SR) and {oscillatory} (L1, L2, IR) phases is of 
{particular} interest 
to spintronics, 
since it emerges from
the interplay of
incoherent spin pumping and  ellipticity. 
{U(1) symmetry is previously taken to be an important condition} in studies of spin-wave lasing~\cite{Bender2012,Bender2014}, {though it is often broken in} magnets with anisotropies. {By explicitly taking this into consideration, our study}
suggests richer phenomena accompanying non-equilibrium phase transitions in spintronic devices.
Although our model description distillates only the essential mechanisms of   an actual spintronics setup, we now briefly discuss some possible implications of our results.  The magnon `lasing'~\cite{Berger1996} in a uniaxial magnet can converge to a steady condensate density featured by a circular precession~\cite{Bender2012,Bender2014}. Turning on interactions explicitly breaking the U(1) symmetry is expected to induce an ellipticity in the spontaneous precession~\cite{Kostylev2012}, accompanied by an oscillation of the condensate density due to the absence of spin conservation. This is similar to the dynamics observed in the   L1 phase   with $\lambda \neq 0$. In the regime where both $\mathbb{Z}_2$ interactions and the pumping effects are sizable, two equivalent $\mathbb{Z}_2$-breaking limit cycles are possible (L2 phase). 
During the electrical pumping, angular momentum transfers reciprocally between the magnet and metal~\cite{Tserkovnyak2014}: as the itinerant electrons exert a spin torque to establish the magnon lasing, the coherent magnetic precession simultaneously pumps a spin current back into the metal~\cite{Tserkovnyak2002}, triggering transverse spin dynamics. Therefore, suppressing the transverse spin dynamics in the metal can be detrimental to magnon lasing, as consistent with the consequence of a fast-relaxing incoherent subsystem discussed above (large $\gamma_t$ limit, see SM for more details).

\begin{figure}[]
\includegraphics[width=1\linewidth]{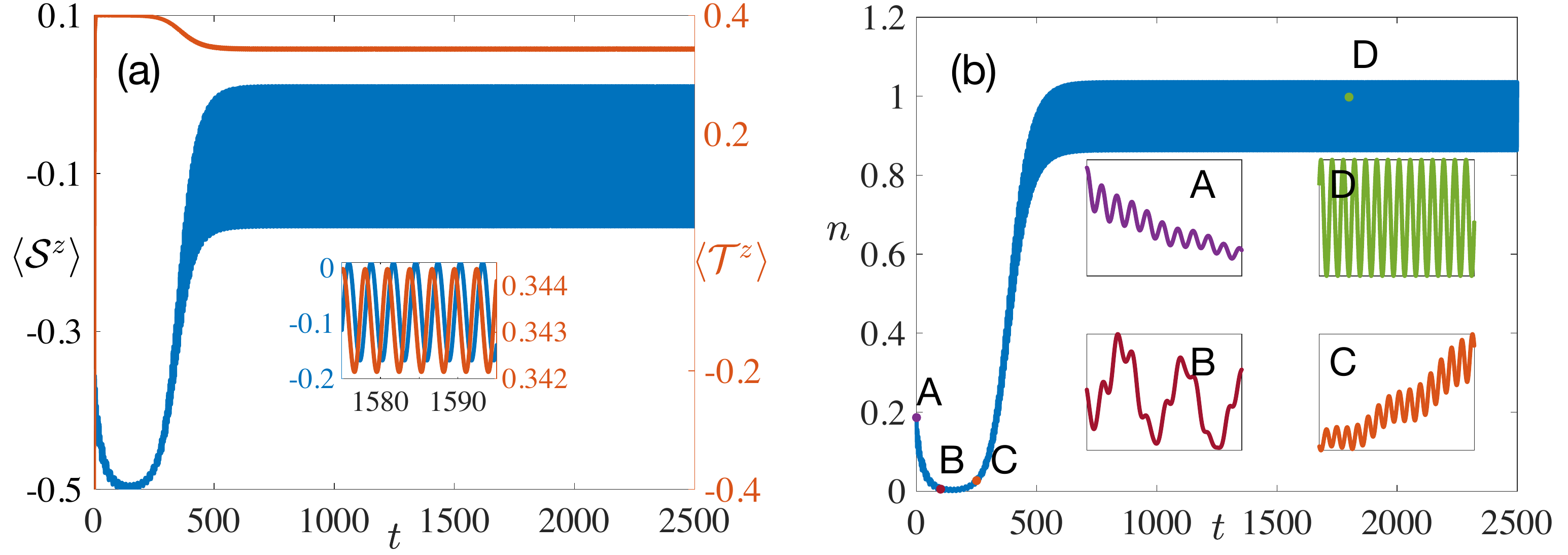}
	\caption{
	{The  dynamics  of  $\langle \mathcal{S}^{z}\rangle$  and $\langle \mathcal{T}^z \rangle$ (a) and the photon number $n$ (b)}  with parameters   as in Fig.~\ref{fig:drawingfigureforpaper},   with the sole exception of   $\omega_z^{\prime}$ that is now chosen    in resonance with the upper polariton frequency ($\omega'_z\simeq\Omega_U$). 
	{Insets show a stretched time axis. }
	The system  is prepared in the SR state and evolved with system's parameters   $\gamma_{\uparrow} $ and $\lambda$ inside the   region L1.   The right panel shows different stages of the photon number evolution. The initial stage of dynamics is governed by the decaying lower polariton mode, while at stage C 
	the upper polaritonic mode undergoes the dynamical instability triggered by the resonance with the incoherent subsystem;    accordingly, the photon number starts to grow until it  saturates around times $t\simeq 500.$ 
 At long times,  $n$ oscillates at the   upper polaritonic frequency (panel D).  This illustrative of  the spin ensemble $\mathcal{T}$  acting as a frequency-dependent gain medium. 
 }
	\label{fig:polar}
\end{figure}

		 \textit{Polaritonic lasing. }  Observables in  the L1 and L2 phases show signatures  of   upper (U) and lower (L) polaritonic modes~\cite{EmaryBrandes}, which
		  are symmetric (U) and anti-symmetric (L) linear superpositions of spin and photon fields, describing   light-matter  hybridization via the Dicke coupling $\lambda$.
		  In order to appreciate this point, we rewrite the
		   interaction term in~\eqref{eq:main_big}  as     $H_{int}=(\eta a+\eta^{\prime}\mathcal{S}^{-})\mathcal{T}^{+} + h.c.$, which is suggestive 
		   {that pumping the $\mathcal{T}$ ensemble can excite a superposition of light and matter in the $\mathcal{S}$ system.}
		   Thus,   upper or lower polaritons can be excited in   the system, depending whether the two couplings have same or opposite sign, jointly with the resonance condition, 
		  $\omega'_z\simeq \Omega_{U / L}$ 
		   (for related expressions, cf. ~\cite{suppmat}). 
The  effective decay   rates of the two polariton modes depend on the  frequency of the incoherent subsystem $\omega_z^{\prime}$ and 
in particular, for  $\eta=\eta^{\prime},$   it can be analytically
		estimated~\cite{suppmat} as 
	{${\kappa}^{\textit{eff}}_{U}$}$=\kappa/2 -2\eta^2 (\gamma_{\uparrow}-\gamma_{\downarrow})/\left[(\omega_z^{\prime}-\Omega_{U})^{ 2}+\gamma_t^2/4\right]$. 
By tuning $\omega_z^{\prime}$ close to $\Omega_{U}$,
where 
$$\Omega_{U / L}=\sqrt{
\left(
\omega_c^2+\omega_z^2 \pm \sqrt{\left(\omega_c^2-\omega_z^2\right)^2+16 \lambda^2 \omega_z \omega_c}\right) / 2}, $$
		 it is possible to obtain a negative effective decay rate (${\kappa}^{\textit{eff}}_U<0$) 
		 for $\gamma_{\uparrow}>\gamma_{\downarrow}$. {The time evolution of $\langle \mathcal{S}^{z}\rangle$, $\langle \mathcal{T}^z \rangle$, and $n$ in this scenario are plotted in Fig.~\ref{fig:polar}. The negative decay rate} gives rise to  dynamical instabilities in the L1 and L2 regimes, which we can 
		{exploit to} employ the spin ensemble $\mathcal{T}$ as   a frequency-dependent gain medium~\cite{gao2015spin,lebreuilly2016towards,PhysRevX.4.031039} (see SM for more details).  
		
		We now discuss  the   multi-stage dynamics [as marked by A to D in Fig.~\ref{fig:polar}(b)] associated with this mechanism. We initialise   the system  in the SR steady state of the Dicke model with photon losses  ({$\lambda = 0.6$ and $\kappa=0.06$}) and  let it  evolve  with parameters characteristic of the L1 phase ({$\lambda=0.2$, $\gamma_{\uparrow}=0.9$}).
		For the chosen parameters, the effective decay rate  {${\kappa}^{\textit{eff}}_L$ (${\kappa}^{\textit{eff}}_U$)} 
		is positive 
		{(negative),  i.e.,}
		the upper polariton becomes unstable.  Immediately after the quench (A), the photon field has sizeable overlap with   the upper and lower polariton modes.   Since in the initial SR steady state the boson is enslaved to matter,  $\langle a\rangle =-2\lambda /(\omega_c-i\kappa/2)\langle \mathcal{S}^{-}\rangle $,  the amplitude of the lower mode 
		is 
		 higher than the amplitude of the upper one.
		 However, as the lower mode starts to decay and the upper one is enhanced,   their amplitudes become comparable (B) and we observe beating at their two frequencies. At the stage (C) the photon number increases 
		 {while} the lower mode is largely suppressed. 
		As a result, for long times (D) the oscillatory dynamics of the system is solely governed by  $\Omega_{U}.$  
		Such circumstance cannot occur in a more conventional  driven-dissipative  Dicke model \cite{Kirton_2018}, since in that case both upper and lower modes would be enhanced and survive at  long times.

	\textit{Outlook.} 	A natural next step  could consist in  studying  collective spin squeezing in the lasing regime~\cite{ma2011quantum,pezze2018quantum,koppenhofer2021dissipative}, with the perspective of entanglement manipulation
    in	spintronics platforms. This can be addressed, for instance, by simulating numerically exact dynamics   at finite   $N$~\cite{PhysRevA.98.063815,lerose2020bridging}.

Recent studies have shown the usefulness of non-local dissipation in generating entanglement between distant qubits in both fields of quantum optics and spintronics, by investigating spins immersed in a optical cavity~\cite{seetharam2021,seetharam2021dynamical,marino2021universality} and nitrogen-vacancy qubits in proximity to a magnetic medium~\cite{zou2021}. For the latter, dynamical phase transitions in the magnet controlled by electrical pumping, may provide an efficient tunability of   non-local dissipation, which could be studied along the lines of this work.
	
Finally,  we did not include here the effect of short-range spin interactions breaking   permutational symmetry. This is in general a challenging task since it     requires a full many-body treatment of dynamics. 
However, we expect that, deep inside the various phases, the dynamical phenomena discussed here will still hold in analogy with the character of other non-equilibrium phases in spin systems with competing short- and all-to-all  interactions~\cite{lerose2019impact,zhu2019dicke}. 

{Our results can be considered as a roadmap to build a novel generation of spintronics experiments inspired by quantum optics, with focus on dynamical phase transitions in heterolayers structures. Scaling up our proof of concept to more concrete platforms appears as an exciting future direction. }\\

		\textit{Acknowledgements.} JM and OC are indebted to P. Kirton for enlightening discussions.	OC  thanks S. Kelly and R. J. Valencia Tortora for helpful comments on this work.  This project has been supported by the Deutsche Forschungsgemeinschaft (DFG, German Research Foundation) – Project-ID 429529648 – TRR 306 QuCoLiMa (”Quantum Cooperativity of Light and Matter”),  by the Dynamics and Topology Centre funded by the State of Rhineland Palatinate, and in part by the National Science Foundation under Grant No. NSF PHY-1748958 (KITP program 'Non-Equilibrium Universality: From Classical to Quantum and Back').
		J.M. and O.C. acknowledge support by the Dynamics and Topology Centre funded by the State of Rhineland Palatinate.
		A.L. acknowledges support by the Swiss National Science Foundation.
		S.Z. and Y.T. are supported by the U.S. Department of Energy, Office of Basic Energy Sciences under Grant No. DE-SC0012190. The Alexander von Humboldt Foundation is acknowledged for supporting YT's stay at Mainz, where this work was initiated.
		I. C. acknowledges financial support from the H2020-FETFLAG-2018-2020 project "PhoQuS" (n.820392), and from the Provincia Autonoma di Trento.

	\bibliography{spintronics}

\newpage\clearpage
		\appendix
\onecolumngrid
	\section{Stability analysis}\label{app:MF}
The mean-field equations of motion used to derive the phase diagram in Fig. 2(a) read    
	\begin{equation}\label{eq:MF}
		\begin{cases}
	\frac{d\langle a \rangle }{d t}=-i \eta \langle  \mathcal{T}^{  -}\rangle -\left(i \omega_{c}+\kappa / 2\right)\langle a \rangle - i \lambda\left(\langle  \mathcal{S}^{+} \rangle +\langle  \mathcal{S}^{-} \rangle \right) \\
\frac{d \langle  \mathcal{S}^{ z} \rangle }{d t}= i \lambda\left(\langle a\rangle +\langle  a^{\dagger} \rangle \right)\left(\langle  \mathcal{S}^{-} \rangle -\langle  \mathcal{S}^{+} \rangle \right)+i \eta^{\prime}\left(\langle  \mathcal{T}^{  +}\rangle  \langle  \mathcal{S}^{-} \rangle -\langle  \mathcal{T}^{  -}\rangle  \langle  \mathcal{S}^{+} \rangle \right) \\
\frac{d \langle  \mathcal{S}^{-} \rangle }{d t}=-i \omega_{z}\langle  \mathcal{S}^{-} \rangle +2 i \lambda\left(\langle a\rangle +\langle  a^{\dagger} \rangle \right) \langle  \mathcal{S}^{ z} \rangle +2 i \eta^{\prime} \langle  \mathcal{T}^{  -}\rangle  \langle  \mathcal{S}^{ z} \rangle \\
\frac{d \langle  \mathcal{T}^{  z}\rangle }{d t}=i \eta\left(\langle  a^{\dagger} \rangle  \langle  \mathcal{T}^{  -}\rangle -\langle a \rangle \langle  \mathcal{T}^{  +}\rangle \right) -i \eta^{\prime}\left(\langle  \mathcal{T}^{  +}\rangle  \langle  \mathcal{S}^{-} \rangle -\langle  \mathcal{T}^{  -}\rangle  \langle  \mathcal{S}^{+} \rangle \right)+\frac{\gamma_{\uparrow}-\gamma_{\downarrow}}{2}-\gamma_{t} \langle  \mathcal{T}^{  z}\rangle  \\
\frac{d \langle  \mathcal{T}^{  -}\rangle }{d t}=-\left(i \omega_{z}^{\prime }+\gamma_{t} / 2\right) \langle  \mathcal{T}^{  -}\rangle +2 i \eta\langle a \rangle  \langle  \mathcal{T}^{  z}\rangle  +2 i \eta^{\prime} \langle  \mathcal{T}^{  z}\rangle  \langle  \mathcal{S}^{-} \rangle.
		\end{cases}
	\end{equation}
	Here we neglected higher order correlations which are all suppressed as $1/N$, approximating $\langle AB\rangle\approx \langle A\rangle\langle B\rangle $. This approximation   is exact in the $N\to\infty$ limit~\cite{kirton2019introduction}. All variables in Eq.~\ref{eq:MF} are intensive, 
	since they are normalized in such a way that they are independent of the number of spins $N$ as $N\to\infty.$
	

	From Eqs.~\eqref{eq:MF} we study the instabilities of the normal state (NS).	
	By  perturbing with small fluctuations around   the normal state expectation values $\langle a\rangle =\langle a\rangle_0 +\delta a, $ $\langle \mathcal{S}^{-}\rangle =\langle \mathcal{S}^{-}\rangle_0+\delta \mathcal{S}^{-}, $ $\langle \mathcal{T}^{  -} \rangle =\langle \mathcal{T}^{  -}\rangle _0+\delta \mathcal{T}^{  -} ,$   (with    
	$\langle a\rangle_0 =\langle \mathcal{S}^{-}\rangle_0=\langle \mathcal{T}^{  -}\rangle _0=0$  and 
	$\langle \mathcal{S}^{z}\rangle =-1/2$ and $\langle \mathcal{T}^{  z}\rangle =(\gamma_{\uparrow}-\gamma_{\downarrow})/2 \gamma_t$), we can find a linear   system of equations for these deviations    from the NS, which can be written in the form $\dot{x}=A x,$
	where $x=\left(  \delta a, \delta a^*, \delta \mathcal{S}^{-}, \delta \mathcal{S}^{+}, \delta \mathcal{T}^{  -}, \delta \mathcal{T}^{  +}\right)^{T}$.
This matrix   reads 
	\begin{equation}\label{eq:mat}
A=		\begin{bmatrix}
			-i\omega_c-\frac{\kappa}{2} & 0 & -i\lambda& -i\lambda& -i\eta & 0\\
			0 & i\omega_c-\frac{\kappa}{2} & i\lambda & i\lambda & 0 & i\eta \\
			-i\lambda & -i\lambda & -i\omega_z & 0 & -i\eta^{\prime} & 0\\
			i\lambda& i\lambda& 0& i\omega_z & 0 & i\eta^{\prime} \\
			2i\eta \langle \mathcal{T}^{  z}\rangle & 0 & 2i\eta^{\prime} \langle \mathcal{T}^{  z}\rangle & 0 & -i\omega_z^{\prime}-\frac{\gamma_t}{2} & 0\\
			0& -2i\eta \langle \mathcal{T}^{  z}\rangle & 0 & -2i \eta^{\prime } \langle \mathcal{T}^{  z}\rangle & 0& i\omega_z^{\prime} -\frac{\gamma_t}{2}
		\end{bmatrix}.
	\end{equation}

\begin{figure}
\begin{minipage}[t]{0.45\linewidth}
	\includegraphics[width=0.9\columnwidth]{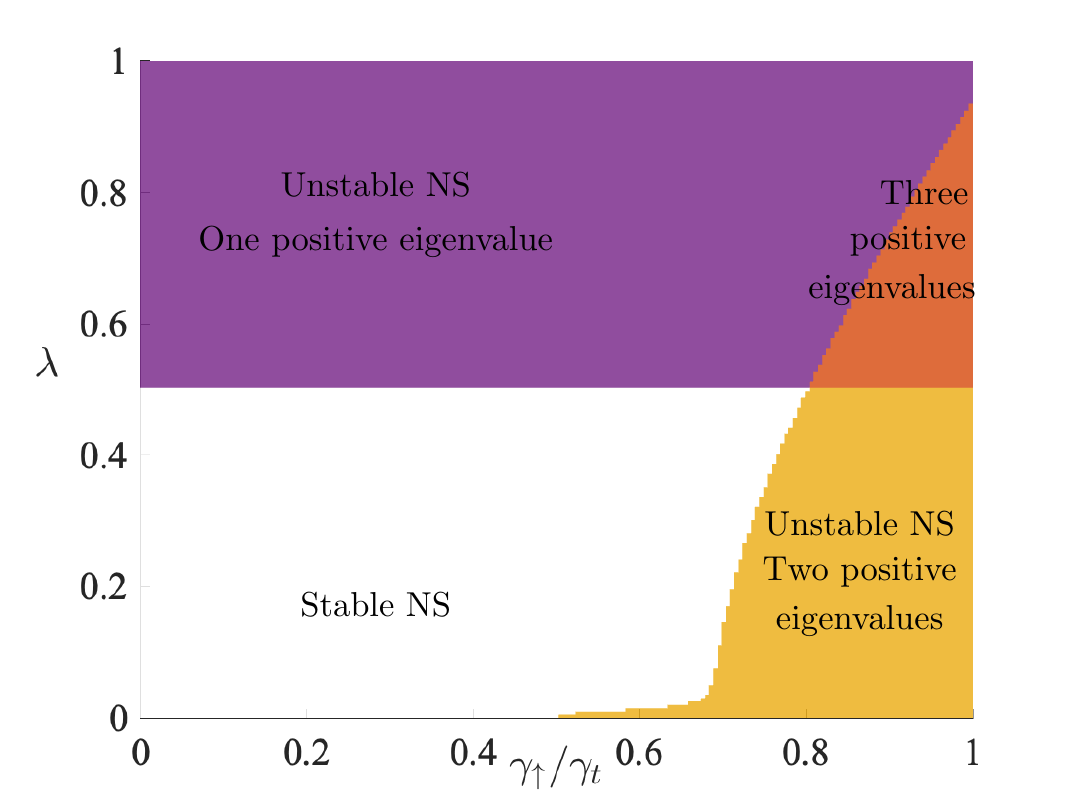}
	\caption{\underline{Stability analysis} of the normal state. The parameters are the same as in Fig.~2(a). The normal state is indicated in white color. The purple color corresponds to one real positive eigenvalue of the matrix $A$ and it indicates  superradiance. The yellow region corresponds to two complex conjugated eigenvalues with positive real part  and it corresponds to the lasing region. The orange region corresponds to the three positive eigenvalues. Here $\omega=\omega_z^{\prime}.$ }
		\label{fig:omegazomegaupuppermode}
\end{minipage}
\hfill
\begin{minipage}[t]{0.45\linewidth}
	\includegraphics[width=0.9\columnwidth]{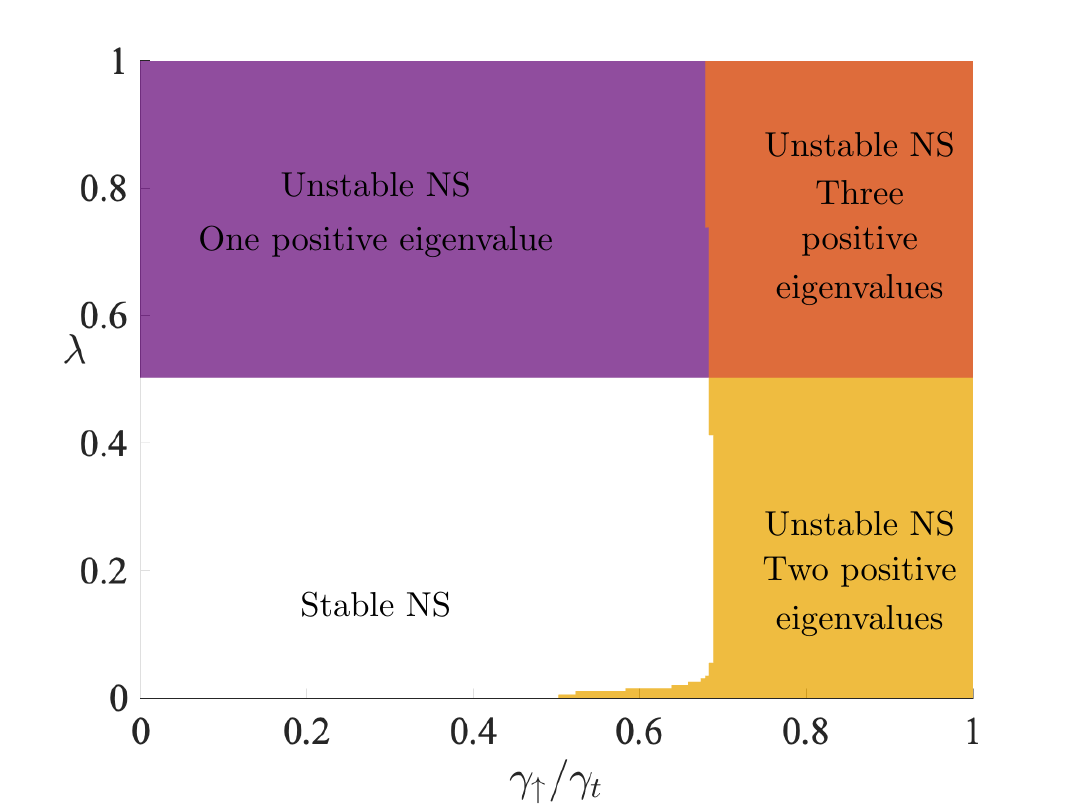}
	\caption{\underline{Stability analysis} with the resonance condition $\omega_z^{\prime}=\Omega_{U}$ . The rest of parameters are   as in Fig.~2(a). The vertical boundary between normal state (white) and lasing (yellow) is given by the critical pumping rate  $\gamma_{\uparrow c}=0.6875$ (see Eq.~\ref{eq:gammac}). }
		\label{fig:omegazomegaupuppermode1}
\end{minipage}
\end{figure}

	By performing  a  stability analysis \cite{Kirton_2018}, one can distinguish the set of parameters for which the normal state is  stable. Results are  shown in Fig.~\ref{fig:omegazomegaupuppermode}. The white region with all negative eigenvalues corresponds to the stable normal phase. The purple region with one real positive eigenvalue matches the   boundary of  SR phase in Fig. 2(a). The yellow region with two positive complex conjugate  eigenvalues corresponds to lasing. The parameters in the orange region corresponds to the three positive eigenvalues of the matrix \eqref{eq:mat}.  The boundary between the SR and NS is well approximated by the $\lambda_c$ of the Dicke model~\cite{EmaryBrandes,Kirton_2018}.
	However, this simple stability analysis does not capture the difference between dynamical phases such as L1, L2 and IR in Fig.~2 (a) of the main text, which require a through evaluation of the far-from-equilibrium dynamics encoded in Eqs.~\eqref{eq:MF}.

	When we pump the system at   a  frequency resonant with the upper polaritonic frequency $\Omega_U$, the boundary of    the lasing region can undergo drastic changes. As it is shown in  Fig.~2, the boundary between normal state and lasing phase is now solely defined   by the critical value of pumping rate $\gamma_{\uparrow}/\gamma_t$, and 
	does not depend on $\lambda.$

	\section{Gilbert damping}\label{app:Gilbert}

 For the magnon condensate mode, dissipation in the form of Gilbert damping slows down the spontaneous precession of the magnetic order parameter like a viscous drag~\cite{Stiles2008}. It is particularly suitable in the weak-damping scenario to describe the relaxation  of the precessional motion back to the equilibrium state (in the absence of external pumping) along a spiral trajectory without losing coherence. The semiclassical Landau-Lifshitz-Gilbert equation~\cite{landau1992,gilbert1955,Gilbert2004} of a spin $\mathbf{s}$ reads $d \mathbf{s}/dt =  \mathbf{s} \times \mathbf{h}_\text{eff} - \alpha_G \mathbf{s} \times d\mathbf{s}/dt$, where $\mathbf{h}_\text{eff}$ is an effective Zeeman field fixing the equilibrium spin orientation, and $\alpha_G$ is the Gilbert damping. 
 A small-angle spin precession can be mapped to the motion of a harmonic oscillator with creation and annhilation operators~\cite{Holstein1940} $a^\dagger$ and $a$, where $s_z = s - a^\dagger a \approx s$ with $\mathcal{S}$ being the spin length. The Landau-Lifshitz-Gilbert equation in the lowest order thus becomes $(1+i s \alpha_G) d \langle a \rangle /dt = -i\omega_c \langle a \rangle$, where $\omega_c = |\mathbf{h}_\text{eff}|.$ Such a form of the viscous drag can also be derived by coupling the bosonic mode $a$ to an Ohmic bath and eliminating the bath degrees of freedom following a standard Caldeira-Leggett derivation \cite{breuer2002theory}.
	\begin{figure}
\begin{minipage}[t]{0.45\linewidth}
	\includegraphics[width=0.9\columnwidth]{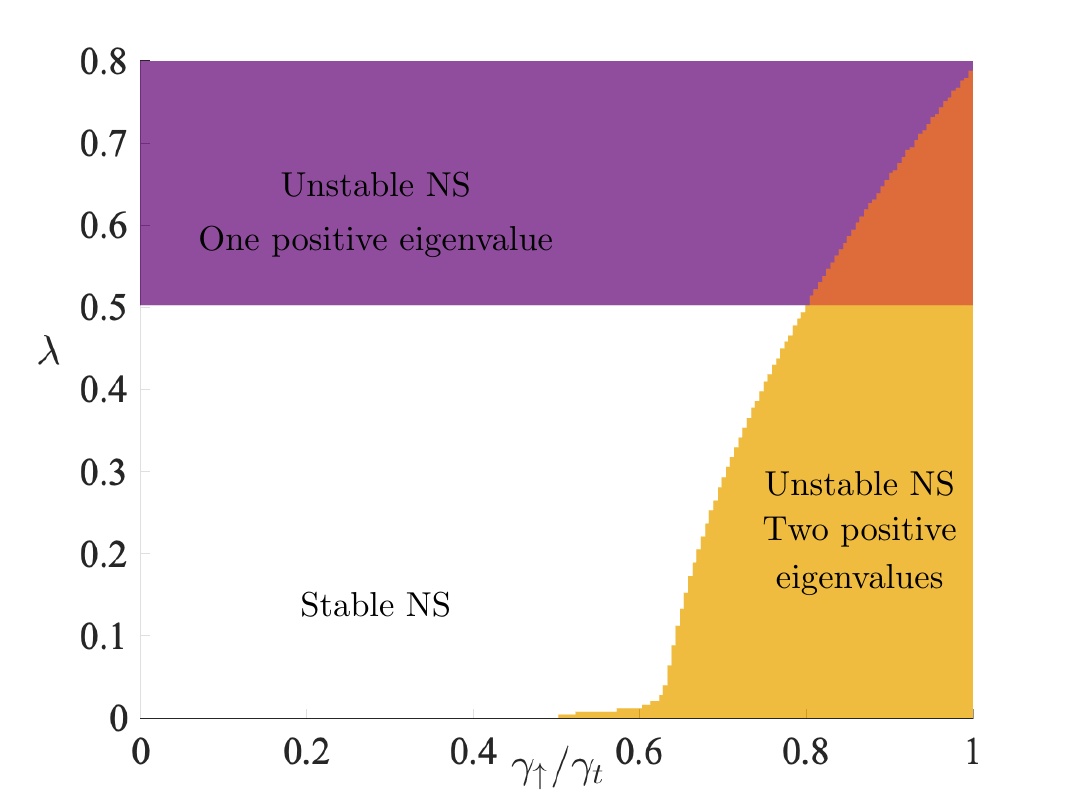}
	\caption{\label{fig:gilbert_SA} \underline{Stability analysis} for model with   Gilbert damping.  All parameters  are as in Fig.~2(a), $\kappa=0.04,$ and the color code follows Fig.~\ref{fig:omegazomegaupuppermode}. }
	\end{minipage}
	\hfill
\begin{minipage}[t]{0.45\linewidth}
	\includegraphics[width=0.82\columnwidth]{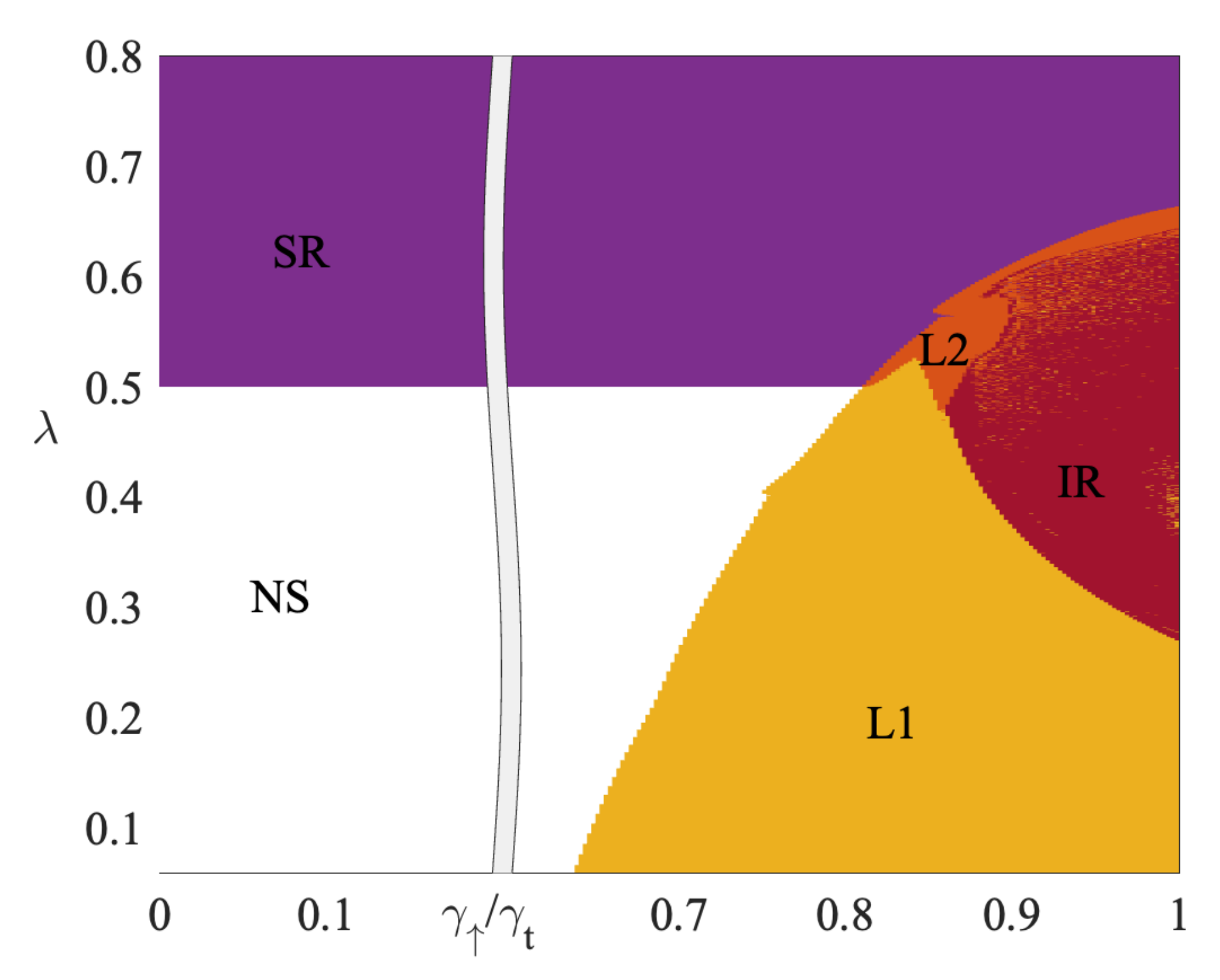}
	\caption{\label{fig:gilbert_phase_diagram} \underline{Dynamical phase diagram} with  Gilbert damping (see Eqn.~\ref{eq:Gilbert}).  All parameters  are as in Fig.~\ref{fig:gilbert_SA} and the color code follows Fig.~2(a).}
\end{minipage}
	\hfill
\end{figure}

For our model (\textcolor{red}{1}),   the Gilbert damping modifies the mean field equation of motion for the expectation value of the bosonic mode into
	\begin{equation}\label{eq:Gilbert}	
		(1+i \kappa/2) \frac{d \langle a \rangle }{d t}=-i\left(\omega_{c}-\tilde{\Delta}\right) \langle a \rangle -i \eta \langle \mathcal{T}^{  -} \rangle - i \lambda\left(\langle \mathcal{S}^{  +} \rangle +\langle \mathcal{S}^{  -} \rangle  \right) +
		\frac{B(t)}{\sqrt{N}}.
	\end{equation}
	Here $\tilde{\Delta}$ is a Lamb shift and  $B(t)$ is the noise term that results from the   bath~\cite{breuer2002theory}. This term is suppressed as $1/\sqrt{N}$ for  large $N$,  therefore vanishing in the mean field limit. 
The dynamical phase diagram with this type of dissipation is plotted  in Fig.~\ref{fig:gilbert_phase_diagram}. Here we fixed $\tilde{\omega}_c=\omega_c-\tilde{\Delta}=1$ and $\kappa=0.04.$ 
Qualitatively, the diagram remains the same as if we used photon losses; we still  recognize five different dynamical responses  as superradiance (SR), normal state (NS), and lasing  and SR persistent oscillations  (L1 and L2), as well as irregular dynamics (IR), although the boundaries between phases are quantitatively modified.  Results are in a good agreement with predictions obtained from the stability analysis (cf. with  Fig.~\ref{fig:gilbert_SA}).

		\section{Lasing in the Tavis-Cummings model}\label{app:Kirton_lasing}
	
 	We consider the limit $\lambda=\eta^{\prime}=0.$   The system in this case is endowed with   a U(1) symmetry~\cite{kirton2019introduction}, corresponding to conservation of total number of excitations (spins+boson);  non-trivial solutions of the mean-field equations of motion \ref{eq:MF} that break dynamically the symmetry can be expressed in the form (see for instance Refs.\cite{tieri2017theory,kopylov2015dissipative})
	\begin{equation}\label{laseso} \langle a\rangle \to a_0 e^{-i\Delta t} ,\quad \langle \mathcal{T}^{  \pm} \rangle  \to  \mathcal{T}_0^{\pm }  e^{\pm i\Delta t}\end{equation} 
  where $\Delta$ is a characteristic frequency to be self-consistently determined.
	Substituting~\eqref{laseso} into Eqs.~\eqref{eq:MF} one   obtains:
	\begin{equation}
		\begin{cases}
			\Delta=\frac{\kappa\omega_z^{\prime }+\gamma_t\omega_c}{\kappa+\gamma_t}\\
			\langle \mathcal{T}^{  z}\rangle =\frac{\gamma_t\kappa}{2 \eta^2}\left( \frac{(\omega_c-\omega_z^{\prime })^2}{(\kappa+\gamma_t)^2}+\frac{1}{4}\right)\\
			\mathcal{T}_0^{\pm }=\sqrt{\frac{\gamma_t \tau_0^{ z}  \kappa}{\eta^2}\left(\frac{(\omega_c-\omega_z^{\prime })^2}{(\kappa+\gamma_t)^2}+\frac{1}{4}
				\right)-\frac{\gamma_t^2\kappa^2}{2\eta^4}\left(\frac{(\omega_c-\omega_z^{\prime })^2}{(\kappa+\gamma_t)^2}+\frac{1}{4}\right)^2 }.\
		\end{cases}
	\end{equation}
	Here $\tau_0^{ z}=(\gamma_{\uparrow}-\gamma_{\downarrow})/2\gamma_t$ is the steady state value of the magnetization in the pumped subsystem. In the $\omega_c\approx\omega_z^{\prime }\equiv\omega$ limit one can further simplify these expressions to acquire physical insight:
	\begin{equation}
		\begin{cases}
			\Delta=\omega\\
			\langle \mathcal{T}^{  z}\rangle =\frac{\gamma_t\kappa}{8 \eta^2} \\
			\langle \mathcal{T}^{  \pm}\rangle =\left(\sqrt{ \frac{\gamma_t \tau_0^{ z}\kappa}{4\eta^2}-\frac{\gamma_t^2\kappa^2}{32\eta^4}    }\right) e^{\pm i\omega t}\\
			n=\frac{\gamma_t}{\kappa}\left(\tau_0^z- \left\langle\mathcal{T}^{z}\right\rangle\right)
		\end{cases}
	\end{equation}
The lasing solution exists only if the expression under square root is positive. This condition is satisfied when the photon loss rate   $\kappa<\kappa_c=4\eta^2/\gamma_t.$   
By introducing photon loses in the form of Gilbert damping (\ref{eq:Gilbert}) and repeating similar calculations, we  find 
\[\frac{2 \eta^{2} \tau_0^{ z}}{\gamma_t }-\kappa / 4 + O(\eta^4)\geq 0.\]
which gives the same critical value of  $\kappa_c.$ 
	
	\section{Instabilities  from adiabatic elimination}\label{app:instability_conditions}

We now work out analytically some dynamical properties of our system in the limit of a fast relaxing bath~\cite{PhysRevA.56.2249}, known as adiabatic elimination  of the bath in quantum optics. 
We   choose $\gamma_t$ large enough compared to $\eta$ and $\eta^{\prime} $ to induce relaxation of the  incoherent subsystem $\mathcal{T}$  much faster than the dynamics of the coherent one ${S}$.
	 Following Refs.~\cite{xu2016theory,norcia2018cavity}, we can enslave the spins of the incoherent ensembles to those of the Dicke system, by setting the time derivatives  of the former to zero:
	\begin{equation}
		\begin{cases}\label{eq:elim}
			\langle\mathcal{T}^{-} \rangle \simeq\dfrac{2 \langle \mathcal{T}^{z}\rangle  (\omega_z^{\prime}+i\gamma_t/2)}{\omega_z^{\prime 2}+\gamma_t^2/4}\left(  {\eta}\langle a\rangle +{\eta^{\prime}} \langle \mathcal{S}^{     -} \rangle  \right)+...,\\
			\langle \mathcal{T}^{+} \rangle \simeq\dfrac{2 \langle \mathcal{T}^{z} \rangle  (\omega_z^{\prime}-i\gamma_t/2)}{\omega_z^{\prime 2}+\gamma_t^2/4}\left(  {\eta}\langle a^{\dagger} \rangle +{\eta^{\prime}} \langle \mathcal{S}^{     +} \rangle  \right)+...\\
			\langle\mathcal{T}{z} \rangle  \simeq(\gamma_{\uparrow}-\gamma_{\downarrow})/2\gamma_t+...,
		\end{cases}
	\end{equation}
	where we neglect terms in higher orders of $1/\gamma_t$.
	This is equivalent to assuming that spins in the $\mathcal{T}$ ensemble have already reached their steady state. 
When   substituting Eq.~\eqref{eq:elim} into the  equations of motion for the normalized  cavity mode and for the spins of the coherent  subsystem $\mathcal{S}$, we find
 
		\begin{equation}\label{cavita}
			\langle \dot{a}\rangle  =-(i\omega_c+\dfrac{\kappa}{2 })\langle  a \rangle -i \eta\left(    \frac{2\eta \langle \mathcal{T}^{   z}\rangle } {(\omega_{z}^{\prime} -\frac{i\gamma_{t}}{2}   )}  \langle  a\rangle +\frac{2 \eta^{\prime} \langle  \mathcal{T}^{   z} \rangle }{ (\omega_{z}^{\prime} -\frac{i\gamma_{t}}{2}   )} \langle  \mathcal{S}^{    -} \rangle  \right) - i \lambda\left(\langle  \mathcal{S}^{    +}\rangle +\langle  \mathcal{S}^{    -}\rangle \right).
		\end{equation}
 	The dissipative dynamics  of the subsystem $\mathcal{S}$ and the photon mode   can now be described with Lindblad terms   with    effective jump operators $L_1=\sqrt{\gamma_{\uparrow}}\tau_{i}^{+}$ and 
 	$L_2=\sqrt{\gamma_{\downarrow}}\tau_{i}^{-}$, given in terms of $a$ and $\mathcal{S}^{-}$ through Eqs.~\eqref{eq:elim}. 
	From Eq.~\eqref{cavita}, we find 
	
	\begin{equation}
		 \langle \dot{a} \rangle =\left(-i\tilde{\omega}-\tilde{\kappa}  \right) \langle a  \rangle - i \lambda\left( \langle \mathcal{S}^{    +} \rangle + \langle \mathcal{S}^{    -} \rangle \right)
		-\frac{2i \eta \eta^{\prime} \langle \mathcal{T}^{   z} \rangle }{ (\omega_{z}^{\prime} -\frac{i\gamma_{t}}{2}   )}  \langle  \mathcal{S}^{    -}  \rangle ,
	\end{equation}
	where 
		\begin{equation}\label{decrate}
	\tilde{\omega}=\omega_c -\frac{2 \eta^{2} \omega_{z}^{\prime} \langle \mathcal{T}^{   z}\rangle}{\left(\omega_{z}^{\prime 2}+\frac{\gamma_{t}^{2}}{4}\right)}, \quad \tilde{\kappa}=\frac{\kappa}{2} - \frac{\gamma_{\uparrow}-\gamma_{\downarrow}}{2} \frac{\eta^{2}}{\omega_{z}^{\prime 2}+\gamma_{t}^{2} / 4}. 	\end{equation}
	According to the second of the  formulas in Eq.~\eqref{decrate}, when the incoherent ensemble is in the population inverted state, the photon mode becomes effectively pumped due to the weak interaction with  $\mathcal{T}$. If this pumping overcomes the photon decay $\kappa$, the photon number starts to grow   and dynamical instabilities are triggered.  

The  equation that effectively governs the dynamics of the coherent subsystem can be derived in the same way and reads   
	\[  \langle \dot{S}^{ -} \rangle =
	 -i \omega_{z}  \langle  \mathcal{S}^{-} \rangle   +2 i \lambda\left( \langle a \rangle +  \langle  a^{\dagger} \rangle  \right)    \langle  \mathcal{S}^{ z} \rangle  
	+\frac{2 i \eta^{\prime}(\omega_z^{\prime}+i\gamma_t/2)(\gamma_{\uparrow}-\gamma_{\downarrow}) \langle \mathcal{S}^{z} \rangle }{\gamma_t(\omega_z^{\prime 2}+\gamma_t^2/4)}(\eta  \langle  a  \rangle + \eta^{\prime}  \langle \mathcal{S}^{-} \rangle ).\]
Here the effective contribution from the dissipator has the form 
\[ \langle \dot{S}^{ -} \rangle \propto \frac{\eta^{\prime }(\gamma_{\uparrow}-\gamma_{\downarrow})(- \langle \mathcal{S}^{z} \rangle ) }{(\omega_z^{\prime 2}+\gamma_t^2/4)}(\eta  \langle  a \rangle + \eta^{\prime}  \langle  \mathcal{S}^{-} \rangle ).\]
	Therefore, for regions with   $\gamma_{\uparrow}> \gamma_{\downarrow}$,    spins in the system are effectively pumped by a rate proportional to the magnetization along $\hat{z}$, provided $ \langle S_z  \rangle $ is negative (as it occurs in the NS or in the SR phase).
	
Adiabatic elimination of the incoherent subsystem gives correct predictions for $\lambda=0$. For $\lambda\neq0$, light and matter hybridize and a separate analysis is required; we elaborate on this  in the next subsection. 

	\section{Polaritons in the dynamics of L1}
	\label{app:polar}

Frequencies of the polaritons can be evaluated by expanding the Dicke model in a leading order Holstein-Primakoff approximation and by diagonalizing the resulting hamiltonian~\cite{EmaryBrandes} 
	\begin{equation}
\Omega_{U/L}=\sqrt{\left(\omega_{c}^{2}+\omega_{z}^{2} \pm \sqrt{\left(\omega_{c}^{2}-\omega_{z}^{2}\right)^{2}+16 \lambda^{2} \omega_{z} \omega_{c}}\right)/2}.
\end{equation}
 For the parameters of  Fig.~\ref{fig:drawingfigureforpaper} the upper polaritonic frequency is
 $\Omega_{U}=1.63$  and the lower one is  $\Omega_{L}=0.63.$ 
 %
 %
 %
  Since the photon amplitude operator $a$ can be 
  written as the sum of upper and lower polaritons, both 
  {modes contribute to}
  the dynamics of $n$. However, their effective decay rates are different, giving rise to different short- and long-time behavior of the the dynamics of $n$.
	%
Inside the SR phase the difference between the amplitudes of upper and lower modes can be estimated using the Holstein-Primakoff analysis~\cite{EmaryBrandes} as
\[\frac{|\psi_{U}|}{|\psi_{L}|}=\left|\frac{1-2\lambda\left(\omega_{c}+i\kappa/2\right)/\left(\omega_{c}^{2}+\kappa^{2}/4\right)}{1+{2\lambda\left(\omega_{c}+i\kappa/2\right)}/{\left(\omega_{c}^{2}+\kappa^{2}/4\right)}}\right|\approx \left|\frac{1-2\lambda/\omega_{c}}{1+2\lambda/\omega_{c}}\right|\]
\noindent which is much more smaller than 1 close to the $\lambda_c.$ 
If we prepare  an initial state inside SR phase and  
let it evolve with parameters
that correspond to any oscillatory phase (L1, L2, IR),
the short-time
dynamics is mostly governed by the lower polariton, as its amplitude dominates.

\begin{figure}
\begin{minipage}[t]{0.45\linewidth}
	\includegraphics[width=0.9\columnwidth]{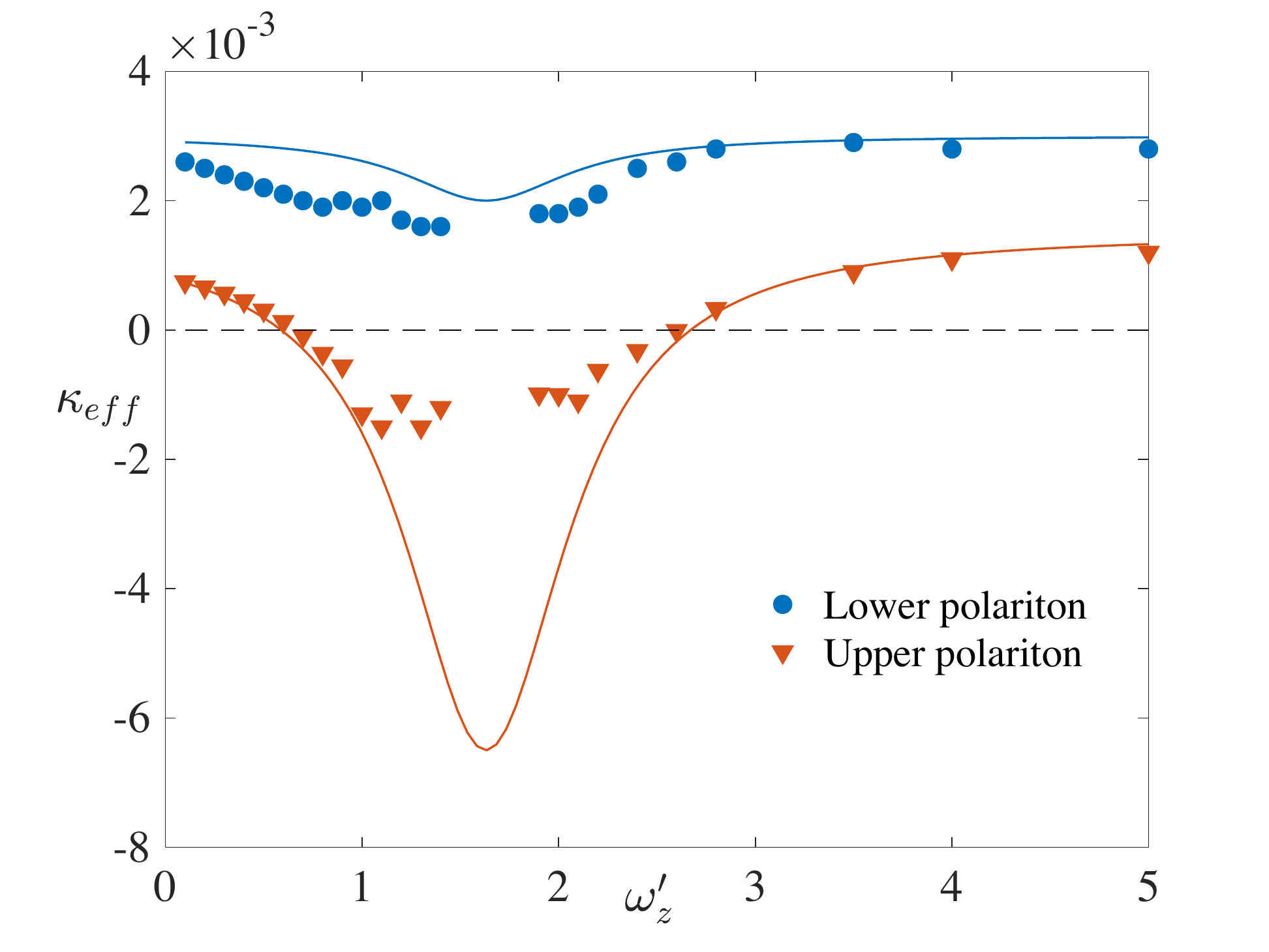}
 		\caption{Effective damping $\kappa_{\textit{eff}} $ of the photonic mode extracted in the  stage C of dynamics in Fig.~3 in the main text, as a function of the frequency of the incoherent subsystem. Close to the resonant frequency of the upper polariton, $\Omega_{U} =1.63$, the effective damping $\kappa_U^{eff}$ can change sign, indicating a dynamical instability, which results into the polariton lasing shown in Fig.~3. 
		\label{fig:resonant-omega}}
\end{minipage}
\hfill	
\begin{minipage}[t]{0.45\linewidth}
	\includegraphics[width=0.9\columnwidth]{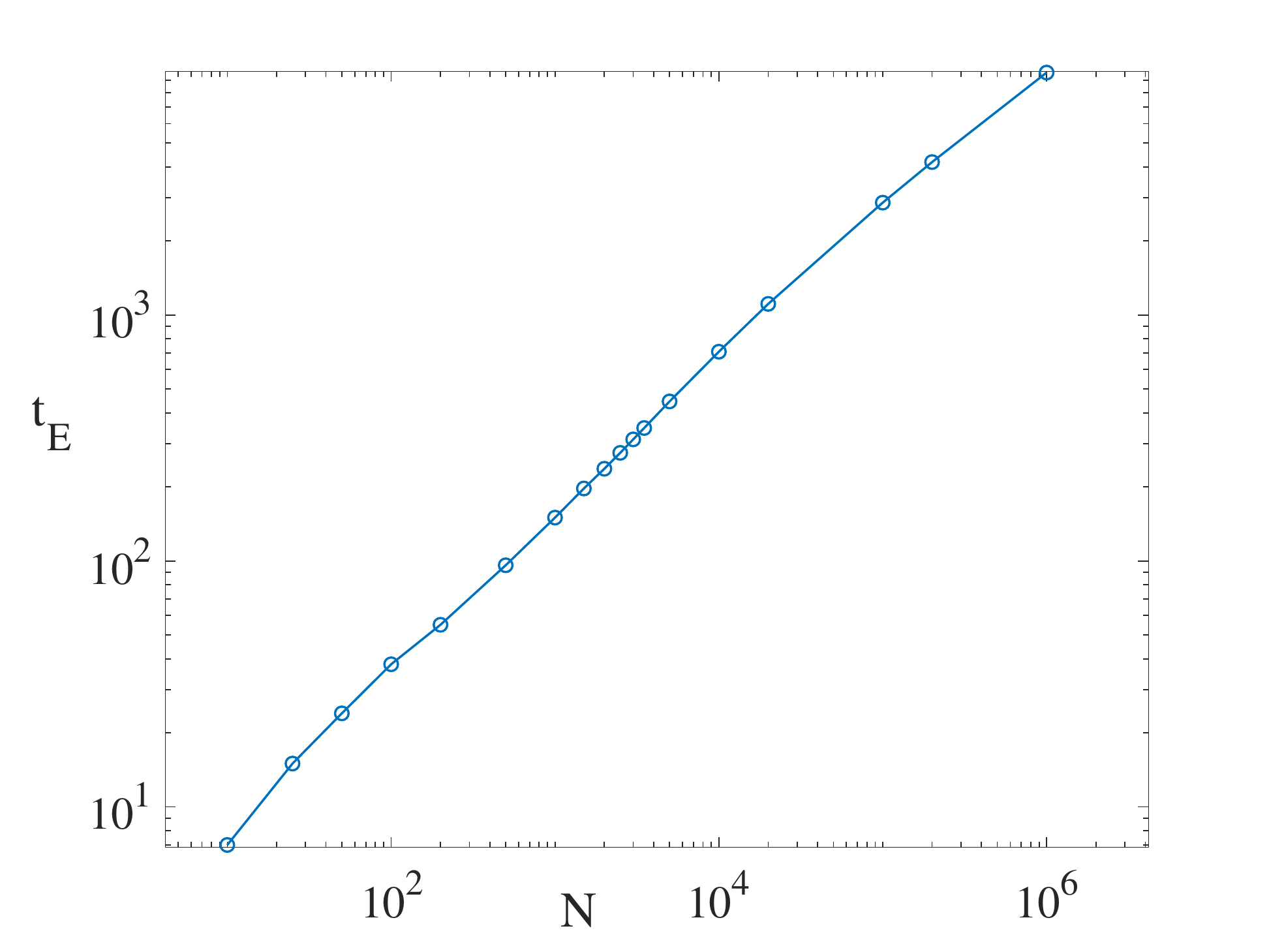}
		\caption{Fit of the  timescale  $t_E\propto N^{\delta}$ with $\delta\simeq0.5$, as a function of number of spins in the system, $N$, within the  lasing region. We extract $t_E$ as the time when the ratio between second cumulants and mean-field expectation values becomes of order $\sim0.1$.}
	\label{fig:loglogerhenfest}
\end{minipage}
\hfill	
\end{figure}

{We now provide estimates for the effective damping of the photon mode in various system's parameters regimes, 
		which is given by 
		$n\propto \exp(-\kappa_{\textit{eff}}~t)$, ignoring oscillations.}
		%
	Depending on the level splitting of the $\mathcal{T}$ spins,  $\omega_z^{\prime}$, 
	$\kappa_{\textit{eff}} $ can be varied. Fig.~\ref{fig:resonant-omega} shows the effective  damping 
	$\kappa_{\textit{eff}}$ of the photonic modes as function of $\omega_z^{\prime}.$  The effective damping coefficients {$\kappa^{\textit{eff}}_{U/L}$}  of lower and upper mode are extracted from dynamics of $n$ at short and long timescales, respectively.
	As one can see from Fig.~\ref{fig:resonant-omega},  
	both damping coefficients have a minimum close to the resonance upper polariton frequency.
	Also, for all frequencies $\omega_z^{\prime}$, the damping of the lower mode is 
{faster}
	than 
	the upper polariton, which is the reason why we observe oscillations at long times with frequency $\Omega_U$ 
	{only}.
	The upper mode is more long-lived 
	and 
	can even be enhanced 
	via pumping,
	when $\omega_z^{\prime} $ is close enough to the upper polaritonic frequency $\Omega_{U}$, 
	resulting in lasing. 
	For $\eta=\eta^\prime$,  the effective damping of
		the 
 		upper 
		polariton mode 
		can be analtyically  estimated as ${\kappa}^\textit{eff}_{U}=\kappa / 2-2 \eta^{2}\left(\gamma_{\uparrow}-\gamma_{\downarrow}\right) /\left[\left(\omega_{z}^{\prime}-\Omega_{U }\right)^{2}+\gamma_{t}^{2} / 4\right]$ (solid 
 		red line 
		in Fig.~\ref{fig:resonant-omega}).

 		If the system is pumped resonantly with the upper  polariton frequency $\omega'_z\simeq\Omega_U$, 
	the critical value of $\gamma_{\uparrow}$ 
{at}
	which the lasing region occurs can be estimated as 
	\begin{equation}\label{eq:gammac}
	    	\frac{\gamma_{\uparrow}}{\gamma_t} \ge \frac{1}{2}\left(\frac{\kappa \gamma_t}{16 \eta^2}+1   \right),
	\end{equation}
following a calculation similar to the one leading to Eq.~\eqref{decrate}. 
	In this case, 
	lasing 
is
	obtained 
	{at a pumping frequency smaller than}
	the conventional threshold for lasing
	$ \omega_z^{\prime}
	\le \Omega_{U} + \sqrt{4\eta^2 (\gamma_{\uparrow}-\gamma_{\downarrow})/\kappa-\gamma_t^2/4}$.

The lower polaritonic mode can be resonantly pumped when $\eta = -\eta^{\prime}$. In this case, the effective damping for both modes have a minimum at the lower polariton frequency $\Omega_L$.
	
	
	\section{Second cumulants}\label{app:secondCumulants}
In models with collective, permutation-symmetric interactions, one can consider   the leading effect of $1/N$ corrections  beyond mean-field, by including  second-order connected correlation functions~\cite{RevModPhys.47.67,haken1984semiclassical}. In general, for finite values of $N$, all higher order connected correlations are relevant for dynamics; however, their effect is expected to be parametrically small in increasing powers of $1/N$ (if $N$ is large). 
This is at the root of the solvability of models with all-to-all interactions mediated by a common bosonic mode, as in our system: the BBGKY hierarchy~\cite{negele2018quantum,huang2009introduction} closes when large system sizes are considered, allowing for non-perturbative solutions   in the couplings governing both unitary or dissipative dynamics. 

We 
include two-point connected correlation functions which couple to mean-field motion,  
neglecting
 third and higher order cumulants by  approximating  three point functions by their disconnected component
\[ \langle ABC\rangle \simeq\langle AB\rangle \langle C\rangle   
+\langle AC\rangle \langle B\rangle  
+\langle BC\rangle \langle A\rangle -2\langle A\rangle \langle B\rangle \langle C\rangle.
\]
We simulate the dynamics  and   compare them with  the  mean-field solution   to  estimate the  timescale,  $t_E$, where cumulants have sufficiently grown to invalidate the mean-field description.
We find that inside the L1 phase $t_E$ scales as the square root of the number of spins (Fig.~\ref{fig:loglogerhenfest}). After   $t_E$, one would have to take into account higher order correlations to correctly  predict the dynamics.  
At times $t\sim \mathcal{O}(N)$ the dynamics of correlators  undergoes phase diffusion~\cite{lewenstein1996quantum, amelio2020theory}.



 \section{Difference between L1 and L2 regions}

\begin{figure}[h]
	    \centering
	   \includegraphics[width=1\linewidth]{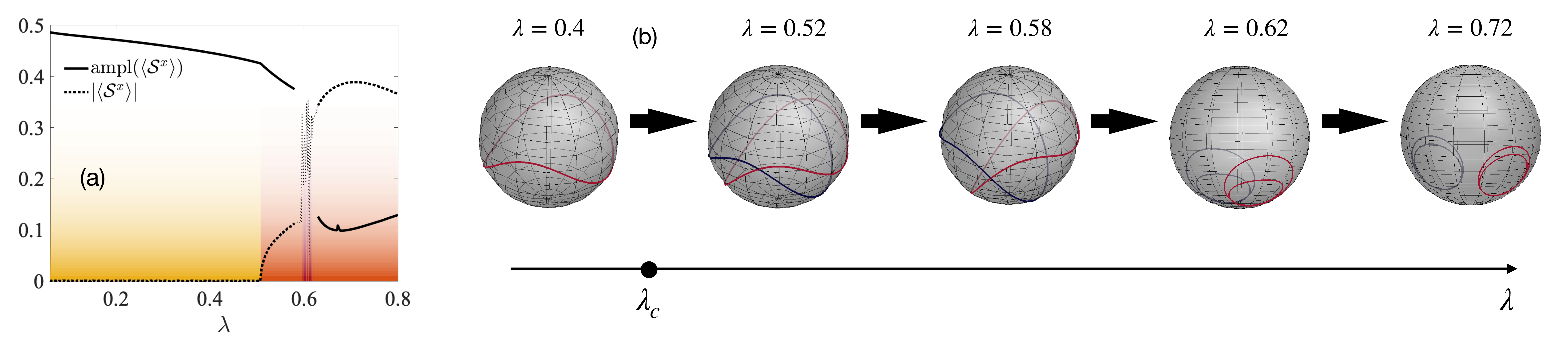}
	    \caption{ (a) Amplitude of the oscillations of the $\langle \mathcal{S}^x\rangle $ (solid line) and absolute value of the time-averaged $\langle \mathcal{S}^x \rangle$ (dashed line) as function of $\lambda.$ We choose all parameters as in Fig.~\ref{fig:drawingfigureforpaper} and fixed $\gamma_{\uparrow}=0.9.$ The colors are the same as in Fig.~\ref{fig:drawingfigureforpaper}. Yellow, orange and red colors correspond to L1, L2 an IR phases, respectively. (b)
	    Dynamics of the spin $\langle \mathcal{S}\rangle $ on the Bloch sphere inside L1 and L2 regions for different values of $\lambda.$ Note that inside L2 region, depending on the initial conditions, one of two trajectories is possible with opposite time-averaged values of  $\langle \mathcal{S}^x\rangle_t=\pm s_0^x.$  }
	    \label{fig:L1L2}
	\end{figure}

	In this Section we consider how the transition between phases L1 and L2 is  captured in dynamics of observables. As we pointed out in the main text, inside L1 region  the dynamics have unbroken $\mathbb{Z}_{2} $ symmetry. Spins components oscillate in time;   the frequency of oscillations of $\langle \mathcal{S}^{z} \rangle $  is twice 
of
	the frequency of oscillations of $\langle \mathcal{S}^x \rangle $ and $\langle \mathcal{S}^y \rangle$. These latter two observables have zero time average. By increasing $\lambda$ above $\lambda_c$, the time-averaged value of $\langle \mathcal{S}^x \rangle$ becomes finite $\langle \mathcal{S}^x\rangle_t=\pm s_0^x$ while the amplitude of oscillations decreases.
	In Fig.~\ref{fig:L1L2}(a) the amplitude of oscillations (solid line) and absolute value of the time averaged $\langle \mathcal{S}^x \rangle$ 
	{(dashed line)}
	are plotted as functions of $\lambda.$
In Fig.~\ref{fig:L1L2}(b) trajectories of $\langle \mathcal{S}\rangle $ for different values of $\lambda$ are shown. Note that for $\lambda>\lambda_c$, depending on initial conditions, one of two trajectories (red or blue lines) are possible with time averaged $\langle \mathcal{S}^x\rangle_t=\pm s_0^x$, respectively.
	%

\end{document}